# Sculpting of lead sulfide nanoparticles by means of acetic acid and dichloroethane

*Frauke Gerdes, Mirjam Volkmann, Thomas Bielewicz,*

*Constanze Schliehe, Christian Klinke\**

Institute of Physical Chemistry, University Hamburg,

Grindelallee 117, 20146 Hamburg, Germany

**Abstract**

*Colloidal lead sulfide is a versatile material with great opportunities to tune the bandgap by electronic confinement and to adapt the optical and electrical properties to the target application. We present a new and simple synthetic route to control size and shape of PbS nanoparticles. Increasing concentrations of explicitly added acetic acid are used to tune the shape of PbS nanoparticles from quasi-spherical particles via octahedrons to six-armed stars. The presence of acetate changes the intrinsic surface energies of the different crystal facets and enables the growth along the <100> direction. Furthermore, the presence of 1,2-dichloroethane alters the reaction kinetics, which results in smaller nanoparticles with a narrower size distribution.*

\* Corresponding author: klinke@chemie.uni-hamburg.de

*Introduction*

The properties of semiconductor nanoparticles (NPs) depend strongly on the particles' size and shape.[1–5] For example, it is possible to tuned the photoluminescence color of CdTe NPs from green to red by increasing their size.[1] Peng *et al.* identified that the Stokes shift between the first absorption maximum and the emission peak is larger in CdSe rods than in dots.[5] To obtain NPs with desired properties, it is essential to understand the nucleation and growth of NPs and to consequently control the growth parameters. Over the last two decades, many approaches have been developed to achieve shape control including using various types of ligands,[6–8] using noble metal seeds,[9,10] and different templates like pores of anodic aluminum oxide membranes or ultrathin tellurium nanowires.[11,12]

Lead sulfide (PbS) is an important IV-VI semiconductor material with a rock-salt structure (cubic face-centred lattice). It has a band gap energy of 0.41 eV and a large exciton Bohr radius of 20 nm.[13] PbS NPs are used in opto-electronic applications such as field-effect transistors,[14,15] optical switches,[13] and solar cells.[16–18] After the first report for a PbS NP synthesis in 2003 by Hines and Scholes[19] PbS NPs with different shapes such as rods, stars, spheres and branched structures have been synthesized using diverse synthetic routes. Cao *et al.* synthesized PbS NPs with different morphologies by using a microwave-assisted method.[20] The hydrolytic synthesis by Pandey *et al.* led to the formation of hexagonal disk, star- and flower-shaped PbS NPs. [21] Zhao *et al.*[22] and Zhou *et al.*[23] achieved shape evolution of PbS NPs in aqueous solution by using thioacetamide (TAA) and cetyltrimethylammonium bromide (CTAB). None of these synthesis routes are based on organic solvents. In contrast, most of the reported PbS NPs synthesized via hot-injection methods are spherical in shape or the shape evolution is obtained by varying several synthesis parameters, like temperature, concentration of the precursor or the type of capping ligands, or by using highly reactive, air-instable or high cost chemicals.[19,24–27] Another way to control the shape is the addition of noble metal particles as seeds to the reaction volume. Li *et al.*[28] used silver seeds to control the growth of PbS NPs to form spheres, cubes, and rods and Yong *et al.*[29] used gold seeds to control the size and shape of PbS nanowires. Thus, simple and inexpensive synthetic routes to control the shape of PbS NPs are desirable.

Here, we present a new synthetic route to control the shape of PbS NPs in a one-pot synthesis. A shape evolution of PbS NPs, from quasi-spherical *via* octahedral to star-shaped nanostructures with six arms along the <100> directions, is simply achieved by explicit addition of various amounts of acetic acid (HAc). A corresponding growth mechanism based on different surface energies of crystal facets and their modulating is proposed to explain this shape evolution. Furthermore, we investigate the influence of 1,2-dichloroethane (DCE) in combination with HAc on the standard synthesis of PbS NPs.

***Results and discussion***

*Synthesis of quasi-spherical PbS nanoparticles*
PbS NPs are synthesized in two steps by the hot-injection method. First, lead acetate is transformed into lead oleate. During this step acetate is removed completely in form of HAc in vacuum. Second, upon heating the lead oleate reacts with TAA to PbS. In the PbS NPs standard synthesis, without additives, quasi-spherical NPs are obtained as shown in Figure 1. The NPs possess an average diameter of 6.0 ± 0.46 nm and a narrow size distribution with a relative standard deviation of 7.7%.

*Influence of acetic acid*
The synthesis of PbS nanoparticles is highly sensitive to additives. Small changes in the type or amount of additive lead to a change in particles shape. The here investigated addition of HAc provides a powerful tool to control the shape of PbS NPs, similar to the case of PbSe NPs.[30] The TEM and HRTEM images in Figure 2 show the different shapes of PbS NPs that are obtained by simply adding HAc to the standard synthesis. By increasing the amount of HAc the particle shape can be tuned from spherical NPs (Figure 2A and 2E; 20 µL HAc) *via* octahedral ones (Figure 2B; 50 µL HAc) to star-shaped ones (Figure 2C and 2F; 100 µL HAc). The HRTEM image in Figure 2F shows the structure of a single star-shaped NP with six symmetric arms in the <100> direction. At even higher amounts of HAc irregular PbS structures are produced (Figure 2D; 150 µL HAc).

X-ray diffraction (XRD) was performed to determine the crystal structure and chemical composition of as prepared samples. As shown in Figure 3 all diffraction peaks can be indexed to the face-centered cubic rock-salt structure of PbS with a lattice constant of *a* = 5.94 Å (comparable to ICCD PDF4 card No. 00-005-0592). Moreover, no further reflexes from impurities such as PbO or $PbS_2$ are observed, indicating that pure PbS NPs are synthesized. In comparison with the diffractogram for bulk and for the standard synthesis, only the octahedral-shaped NPs show small changes of the intensity ratio of the (111) reflex to the (200) reflex (Figure 3B). The higher intensity of the (111) reflex suggests that the growth takes place along the <100> directions. The quasi-spherical NPs (Figure 3A) do not show differences, because the shape of the NPs does not change by the addition of 20 µL HAc. For the star-shaped (Figure 3C) and the irregular-structured particles (Figure 3D) no deviation from the standard synthesis can be observed since those structures are already pretty large.

In addition to the variation of the shape the HAc has also an effect on the size and the relative standard deviation of the NPs. With increasing amount of HAc the size of the PbS NPs increases. The quasi-spherical PbS NPs, which are synthesized with 20 µL HAc, possess an average diameter of 7.3 ± 0.74 nm with a relative standard deviation of 10.2%. For the octahedrons an average diameter of 10.4 ± 2.1 nm with a relative standard deviation of 20.1% is observed. The PbS stars have a particle size of about 37 nm (tip to tip).

The effect of increased size by increasing the amount of HAc can be also observed in the XRD patterns (Figure 3). A larger width of the XRD reflexes indicates a reduced particles size. The average size of the crystalline domains of the NPs can be determined from the full width half maximum of the reflexes by using the Scherrer equation.[31] The NPs of the standard synthesis have a size of 5.6 nm, while the NPs grown with 20 µL have a size of 5.7 nm. For the octahedrons, stars and unregulated structures following average sizes have been determined: 10.1 nm, 18.0 nm and 16.9 nm respectively. The result of the evaluation of the (220) reflex is that the calculated sizes from the XRD are smaller than the one from the TEM evaluation. The reason is that XRD reflects the crystalline parts and in the TEM the particles might appear a bit larger than only the crystalline part. Additionally, for the anisotropic octahedrons and stars the XRD size averages over all directions giving a smaller number than the length of the extensions.

The absorption spectra of the standard synthesis and of the quasi-spherical NPs (20 µL HAc) show clear first absorption maxima (Figure 4). Since the 20 µL particles are a bit larger the absorption maximum shifts to longer wavelengths. For the other structures, octahedral and star-shaped PbS NPs, no absorption maximum was detected in the measured range.

*Growth mechanism*

The nanoparticle formation can be separated into two phases: the nucleation and the growth process. During the nucleation the NP nuclei form a specific crystalline phase, but the final geometry of NPs will be determined by the growth rate of the different crystal facets. The growth rates in different crystal directions can be tailored by modulating the relative surface energies of those crystal planes by using surface selective ligands. The rock-salt PbS NPs nucleate as tetradecahedron seeds, which expose six {100} facets and eight {111} facets.[32,33] Thereby the {111} facets are terminated by either Pb atoms or S atoms, while the {100} facet consists alternating of Pb atoms and S atoms. In the absence of capping ligands, the surface energy of the {111} facets is higher than that of the {100} facets and the faster growth along the equivalent {111} facets will result in the formation of cube-shaped NPs.[22,32]

In this work, we use two different ligands to passivate the particle surface, oleate and acetate. When the PbS NPs are synthesized using the standard synthesis conditions containing only oleate as ligands, quasi-spherical PbS NPs were obtained. The additional use of HAc as a further capping molecule leads to an evolution of the PbS NP shape. Although both molecules have a negatively charged carboxyl functional group, they have different effects on the modulation of the surface energies because of their different length. Choi *et al.*[34] calculated, that acetate-capped Pb-rich {111} surfaces have lower surface energies than the {100} surfaces, while oleate-capped Pb-rich {111} surfaces exhibit higher surfaces energies. The different steric hindrance between the molecules is responsible for the particular effect on the surface energies.

With increasing HAc volume the shape and the size of the PbS NPs change significantly. The addition of HAc to the reaction mixture leads to a partial exchange of oleate against acetate at the Pb-rich facets on the NPs surface, because the acetate molecules are much smaller than oleate and the steric hindrance is strongly reduced. With increased number of acetate molecules on the {111} facets, the surface energy of the {111} facets decrease relative to the {100} facets until it is lower and the growth along the <100> directions is faster.

When the HAc volume is increased to 20 µL no significant changes appear in the shape, but the size increase from to 6.0 nm (0 µL HAc) to 7.3 nm. The added amount of HAc is not enough to lower the relative surface energy of the {111} facets to that of the {100} facets so that the growth along the <100> direction is not favored. When the amount of HAc is further increased (50 to 100 µL) the acetate on the particle surface leads to a change of the growth rate and the growth along the <100> direction results in octahedral-shaped (10.4 nm) and six-armed star-shaped (~ 37 nm) PbS NPs. By further increase of the volume of HAc irregular-structured particles are formed. The increasing number of acetate molecules on the surface of the NPs finally leads to a reduced protection and to a destabilization of the particles, which results in agglomeration.

*Combined influence of HAc and DCE*
Halogenated molecules are used to perform ligand exchange reaction to efficiently displace other ligands and to improve the passivation of the NPs. [35,36] Chlorine-containing compounds can be also used as etching agents, altering the initial shape of NCs, [37] or as complexing agents to manipulate the reaction kinetics. [38]

In earlier works, we showed that chlorine containing additives can have a strong influence on different NP synthesis: CdSe nanorods could be transformed into hexagonal pyramids by halogenated additives such as DCE.[39] We found that at synthesis temperatures around 250 °C haloalkanes work very similar to ionic halogen additives and coordinate to Cd-rich facets. This leads to a modified chemical composition of the ligand sphere and influences both the kinetic and thermodynamic growth regime to achieve the shape transformation. DCE also is essential for the formation of two-dimensional PbS nanosheets.[40] It acts as lead complexing agent and alters the kinetics of nucleation and growth. As a result, small NPs are formed, which exhibit highly reactive {110} facets supporting an oriented-attachment process. In both cases, as L type (coordinative) or X type (ionic) ligand, DCE is responsible for changes in the reaction kinetics and consequently for the shape alteration.

In the present work, DCE is added to the reaction mixture simultaneously with the HAc to investigate the combined influence of both additives to the PbS standard synthesis. The TEM images show the same shape evolution compared to the syntheses without DCE (Figure 5). The DCE has no strong impact on the shape of the PbS NPs under these conditions; but DCE influences other aspects of the NPs. In the presence of DCE smaller NPs are formed with a narrower size distribution. For example, when adding 20 µL HAc without DCE 7.3 nm sized quasi-spherical particles with a standard deviation of 10.2% are formed, while in the presence of DCE 6.1 nm sized particles with a standard deviation of 8.8% are observed. For the octahedral-shaped NPs the difference of the standard deviation is more pronounced. The NPs without DCE show a standard deviation of 20.1%, whereas the NPs with DCE only have a standard deviation of 8.9%.

DCE which acts as a lead complexing agent might alter the kinetics of nucleation and growth of the NPs. A similar trend was already observed by Schliehe *et al.*[40,41] and Reilly *et al.*[40,41] The complexation of the lead ions leads to a nucleation at higher supersaturation, whereby more nuclei are formed and this results in smaller NPs, according to the LaMer model.[42] The reduced size of the NPs is also favored by a strong passivation of the NPs. The additional ligand DCE might lead to a stronger passivation of the surface, whereby the growth rate could be lowered and smaller NPs are formed. This decreased reaction rate can also be observe in the later color change of the solution during the course of the reaction, which indicates the PbS NPs formation. In comparison to the synthesis with only HAc, the color change occurred one to two minutes later using DCE.

The narrower size distribution of the smaller NPs is caused by the higher solute concentration after the nucleation. The following diffusion-controlled growth makes smaller NPs grow faster than larger ones, which results in a more narrow size distribution.

*Conclusion*

In summary, we presented a new and simple path to synthesize monodisperse PbS NPs, in which lead oleate and thioacetamide were used as reactants, and oleate and acetate were used as capping ligands. The NPs exhibit different shapes due to a modulation of the surface energy and the growth rate of the {100} and {111} facets by different amounts of HAc. The replacement of oleate against acetate led to a faster growth along the <100> direction and the formation of quasi-spherical, octahedral and six-armed star-shaped particles could be observed. The presence of DCE leads to smaller and more monodisperse NPs.

*Experimentals*

*Materials*
Lead(II) acetate trihydrate (99.9%), *N,N*-dimethylforamide (≥ 99%), diphenyl ether (99%), acetic acid (≥ 99.7%), oleic acid (90%), thioacetamide (≥ 99.0 %) and toluene (≥ 99.7%) were purchased from Sigma-Aldrich. Acetonitrile (99.9%) was ordered from Acros Organics, and Ethanol (p.a.) and 1,2-dichloroethane (≥ 98%) were obtained from Merck. All chemicals were used without further purification.

*Methods*

Synthesis and purification of PbS NPs

The S precursor solution was prepared by dissolving thioacetamide (0.08 g, 1.1 mmol) in a mixture of *N,N*-dimethylforamide (0.5 mL) and diphenyl ether (6.0 mL) and stored in a nitrogen filled glove box.

A mixture of lead(II) acetate trihydrate (0.87 g, 2.3 mmol), oleic acid (3.5 mL, 11 mmol) and diphenyl ether (15 mL) was heated to 85 °C to form an optically clear solution. To remove acetate, which is formed during the conversion of lead acetate into the oleate, the solution was stirred for 1.5 h under vacuum at 85 °C. Afterwards the temperature was raised to 100 °C and 1 mL (0.16 mmol TAA) of the S precursor solution was injected under nitrogen atmosphere. The color of the solution turned from colorless to dark brown, indicating the formation of PbS NPs. The reaction was quenched after 10 min by cooling down to room temperature.

In order to investigate the influence of HAc and the combination with a chlorine-containing additive on the PbS NPs synthesis, different volumes of HAc (20 – 150 µL) and a constant volume of DCE (1.0 mL, 12.6 mmol) were added to the reaction solution after the *in situ* formation of lead oleate at 85 °C. After 10 min the temperature was raised to 100 °C.

All resulting PbS NPs were purified by precipitation with acetonitrile, centrifugation (4500 rpm, 5 min), removal of the supernatant and re-suspension in toluene for three times to remove any residuals.

*Characterization*

The morphology of the PbS NPs was analyzed by transmission electron microscopy (TEM) and high-resolution TEM (HRTEM). TEM images were performed on a *JEOL Jem-1011* microscope at an acceleration voltage of 100 kV. A *Philips CM 300* microscope with an acceleration voltage of 200 kV was used for HRTEM. The samples were prepared by dropping dilute toluene solutions of PbS NPs onto carbon-covered copper grids.

The composition of the PbS NPs was determined by X-ray diffraction (XRD) measurements, which were carried out on a *Philips X'Pert PRO MPD* diffractometer with monochromatic X-Ray radiation from a copper anode with a wavelength of 1.54 Å ($CuK_\alpha$). The samples were prepared by dropping toluene solutions of PbS NPs onto a silicon wafer.

The absorption measurements were carried out in a quartz vessel with an optical path length of 10 mm with a *Varian Carry 5000* two-beam spectrometer. For the measurements the samples were transferred into 1,1,2,2-tetrachloroethane by two cycles of precipitation with acetonitrile and re-dispersion with the new solvent.

**Acknowledgments**

The authors thank the European Research Council (Seventh Framework Program FP7, Project: ERC Starting Grant 2D-SYNETRA) for funding. CK acknowledges the German Research Foundation DFG for a Heisenberg scholarship.

**Figures**

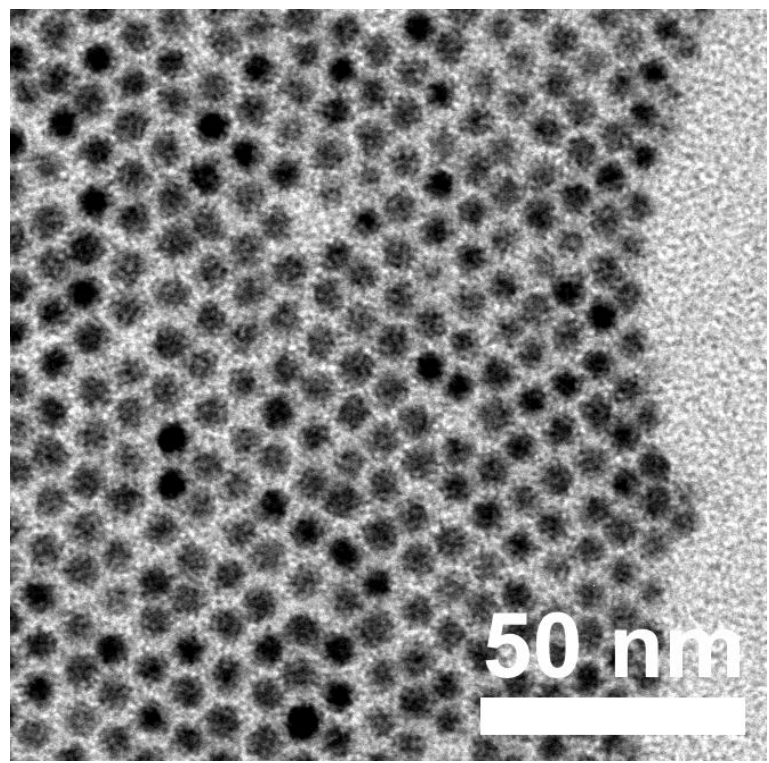

Figure 1: *TEM image of quasi-spherical PbS NPs synthesized at standard conditions (without HAc or DCE).*

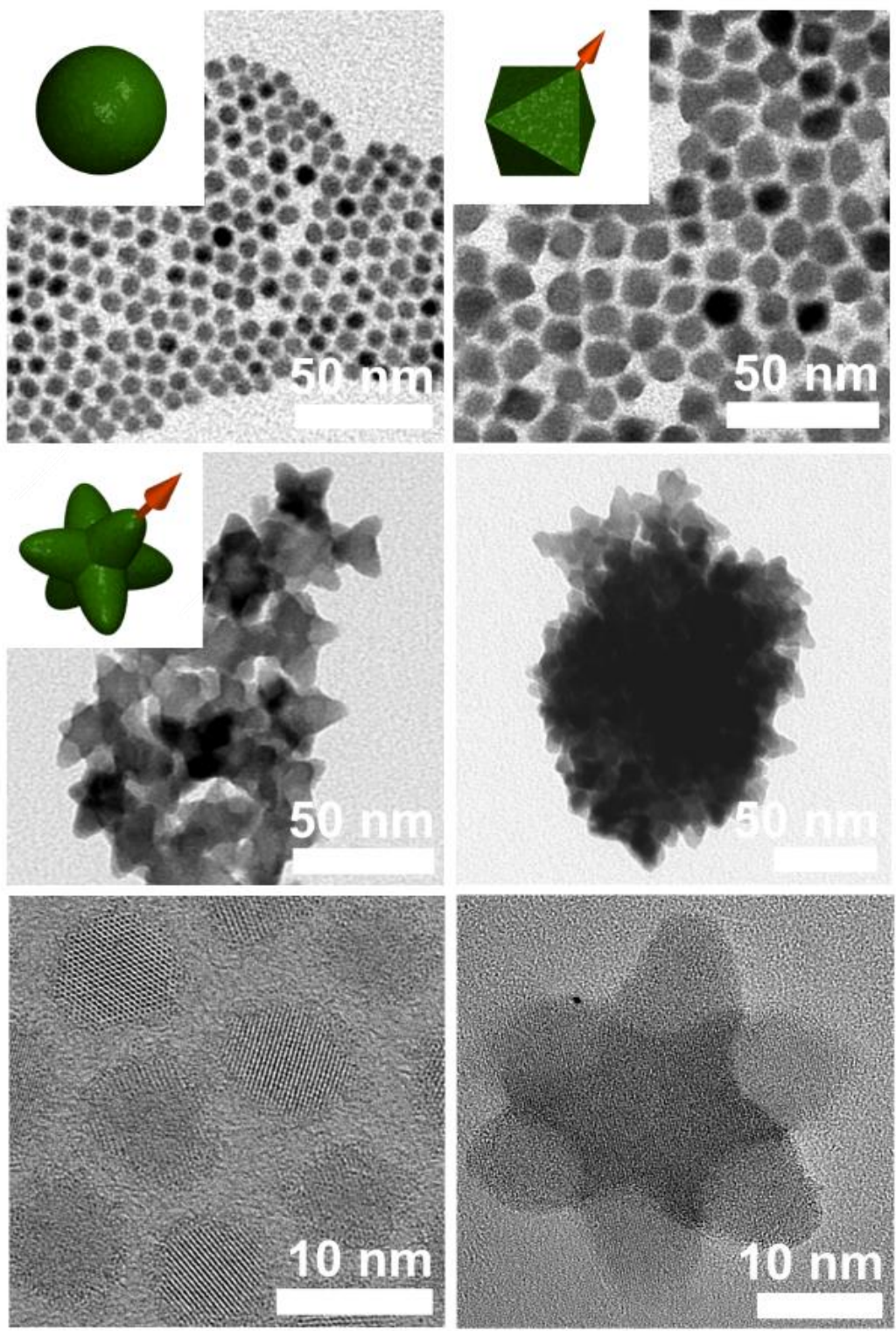

Figure 2: *(A) – (D) TEM images of PbS NPs of different shapes grown with different volumes of HAc. (A) quasi-spherical (20 µL HAc), (B) octahedral (50 µL HAc), (C) star-shaped (100 µL HAc) and (D) irregular-structured (150 µL HAc) particles. (E) – (F) HRTEM images of (E) quasi-spherical (20 µL HAc) and (F) six-armed star-shaped (100 µL HAc) PbS NPs. The insets are models of the structures. The orange arrow indicates the <100> direction.*

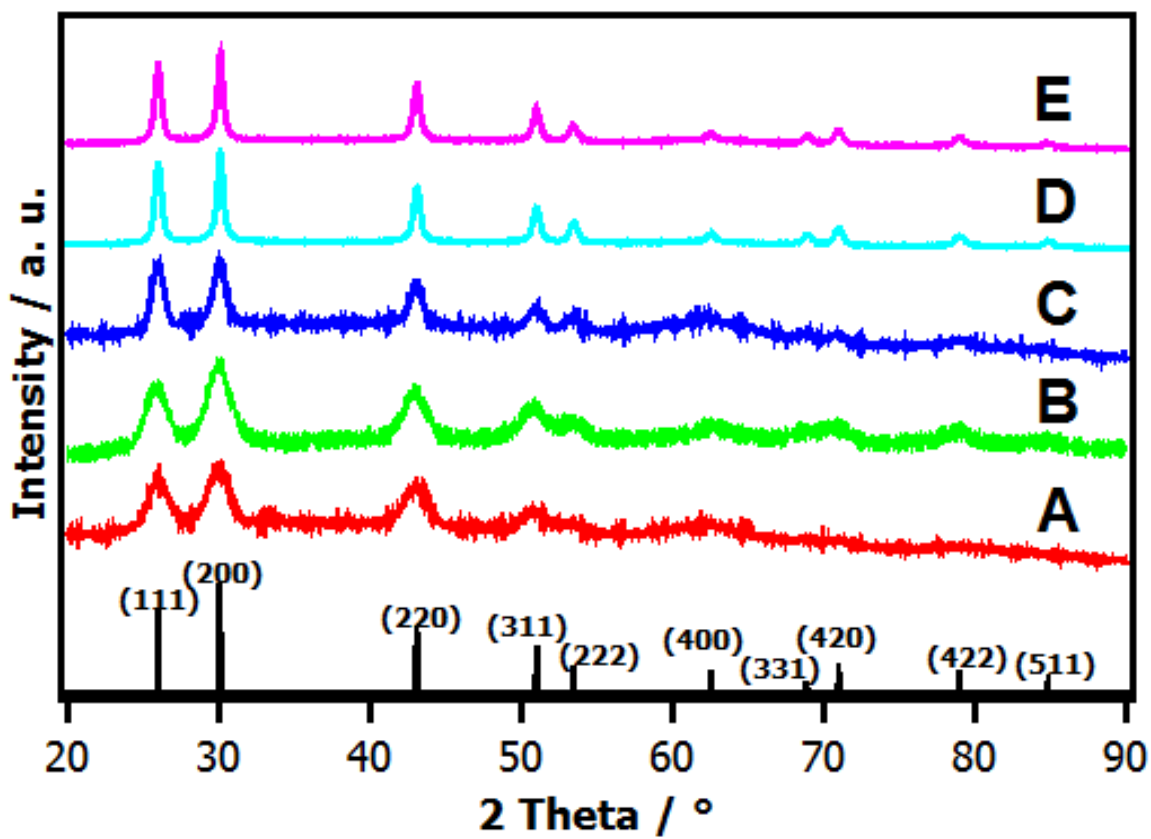

Figure 3: *Powder X-ray diffractograms of PbS NPs with different shapes grown with different volumes of HAc: (A) spherical (0 µL HAc), (B) quasi-spherical (20 µL HAc), (C) octahedral (50 µL HAc), (D) star-shaped (100 µL HAc), and (E) irregular-structured (150 µL HAc) particles. At the bottom the diffractogram of bulk PbS is shown (ICCD PDF4 card No. 00-005-0592).*

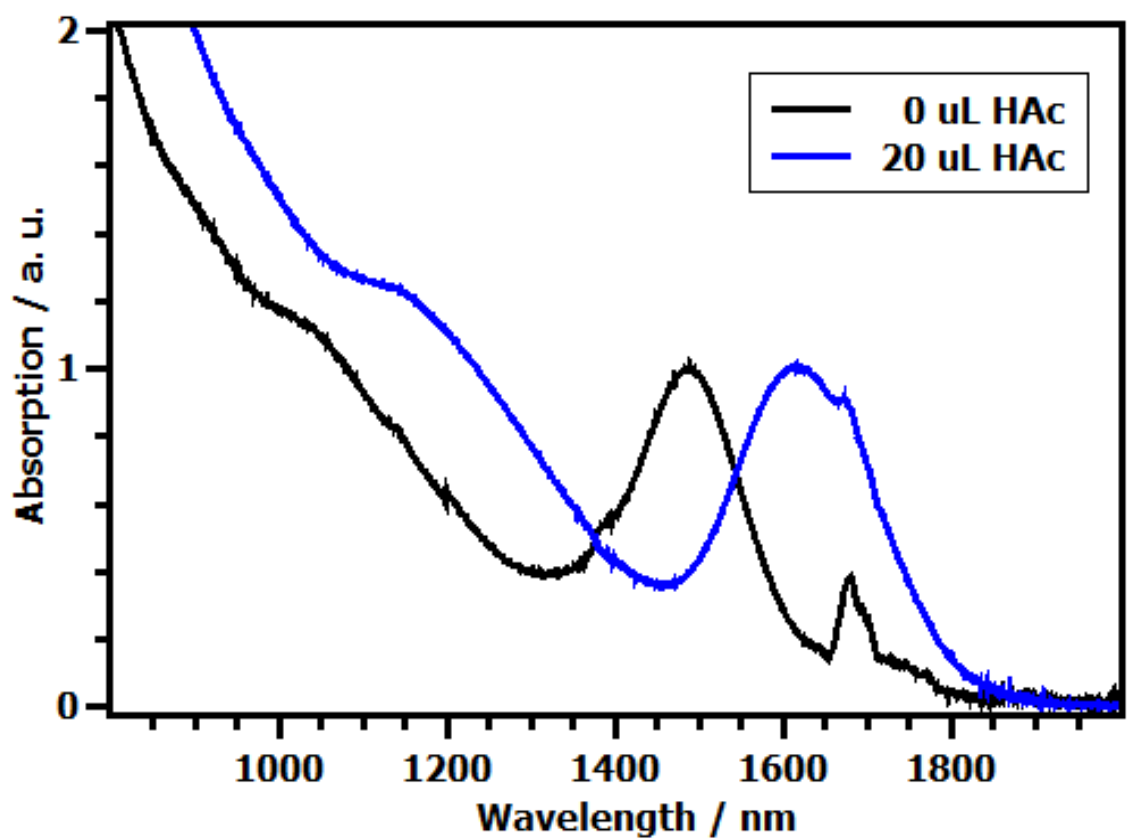

Figure 4: *Absorption spectra of PbS NPs without HAc (black) and with 20 µL HAc (blue). The other shapes do not show clear absorption maxima in the measured range.*

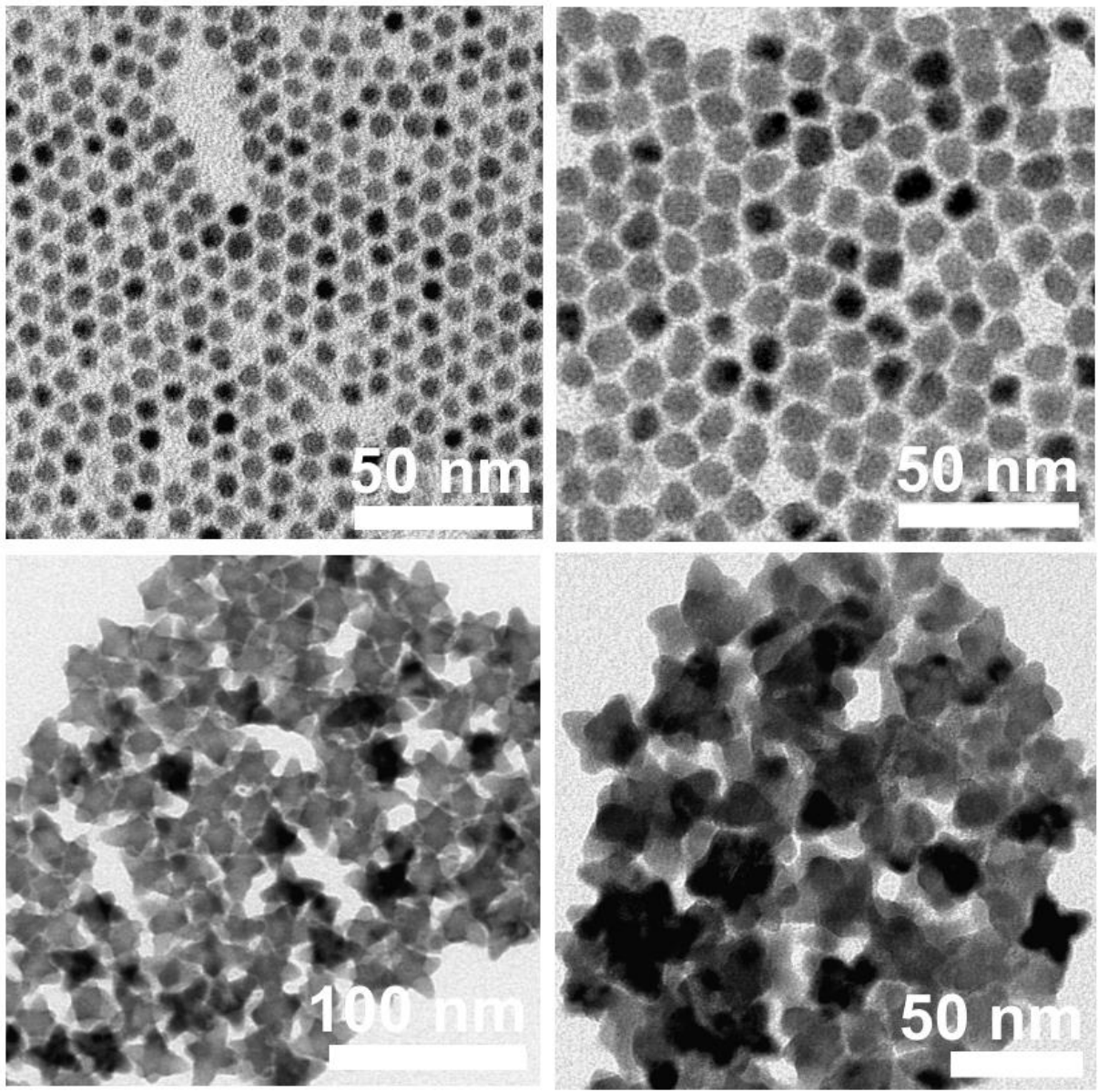

Figure 5: *TEM images of PbS NPs with different shapes grown with different volumes of HAc in the presence of DCE. (A) quasi-spherical (20 µL HAc), (B) octahedral (50 µL HAc), (C) star-shaped (100 µL HAc) and (D) irregular-structured (150 µL HAc) particles.*